\documentclass{PoS}

\title{Dark Matter Annihilation Signals:\\ The Importance of Radiative Corrections}

\ShortTitle{Radiative corrections to DM annihilation}

\author{\speaker{Torsten Bringmann}
\\
        Department of Physics, Stockholm University, AlbaNova
 University Center, SE - 106 91 Stockholm, Sweden\\
        E-mail: \email{troms@physto.se}}

\abstract{Being able to safely distinguish astrophysical from potential dark matter (DM) annihilation signals is of utmost importance for indirect DM searches. To this end, one has to rely on distinctive -- and unique -- spectral signatures to look for. Internal bremsstrahlung (IB), unavoidable in the presence of charged annihilation products, provides such a signature. In fact, as it \emph{generically} dominates the gamma-ray spectrum expected from DM annihilations, at high energies, it may well turn out to be more important for indirect DM searches than the traditionally looked-for line signals. As illustrated in some detail, the observation of IB signatures would even allow to distinguish between different DM candidates or to constrain significantly the parameter space of, e.g., neutralino DM.
The gamma-ray contributions reported here are therefore of great interest for the already launched Fermi/GLAST satellite and the upcoming new generation of Air Cherenkov Telescope systems like CTA -- which are most sensitive at the high energies where these effects are particularly important. Finally, radiative corrections may even significantly alter the positron spectrum from DM annihilations; an intriguing positron excess recently found by the PAMELA satellite might turn out to be an indication of the peculiar spectral signature expected in that case.
}

\FullConference{Identification of dark matter 2008\\
		 August 18-22, 2008\\
		 Stockholm, Sweden}

\begin{document}

\section{Introduction}

Evidence for the existence of a sizeable contribution of non-baryonic, cold dark matter (DM) to the total matter content of the universe has in recent years become overwhelming, with a wealth of independent observations consistently pointing towards a cosmological concordance model that is fully described by only a handful of parameters. While the nature of this mysterious DM component still remains unresolved, 70 years after the first hints about its existence,  it may actually be regarded as the first direct evidence for physics beyond the standard model of particle physics (SM). In fact, weakly interacting massive particles (WIMPs), that appear in most SM extensions at the electroweak scale (obviously of considerable interest to the soon-to-start LHC at CERN), are particularly well-suited DM candidates: thermally produced in the early universe, they naturally provide the right relic density to account for the DM abundance observed today. The theoretically maybe best motivated, and certainly most studied, DM candidate in this context is the supersymmetric neutralino. For reviews of WIMP DM see, e.g., \cite{review}.

Searches for this type of particle DM can be grouped in three categories: \emph{indirect searches} for SM particles created by DM annihilation in the galactic halo (or, in the case of neutrinos, in celestial bodies like the sun or earth), \emph{direct searches} for DM particles recoiling from the nuclei of large terrestial detectors or \emph{accelerator searches} that look for missing transverse energy. In all of these cases, the signal is likely to be very weak and/or dominated by a much larger background. Clear signatures that allow a conclusive identification of an event as being DM induced are thus mandatory. In this contribution, I will focus on indirect detection methods and point out that radiative corrections to the leading order annihilation processes provide such signatures.

\section{Gamma rays}

\begin{figure}[t]
\vspace*{-0.5cm}
\begin{minipage}[t]{0.49\textwidth}
\centering
\includegraphics[width=\textwidth]{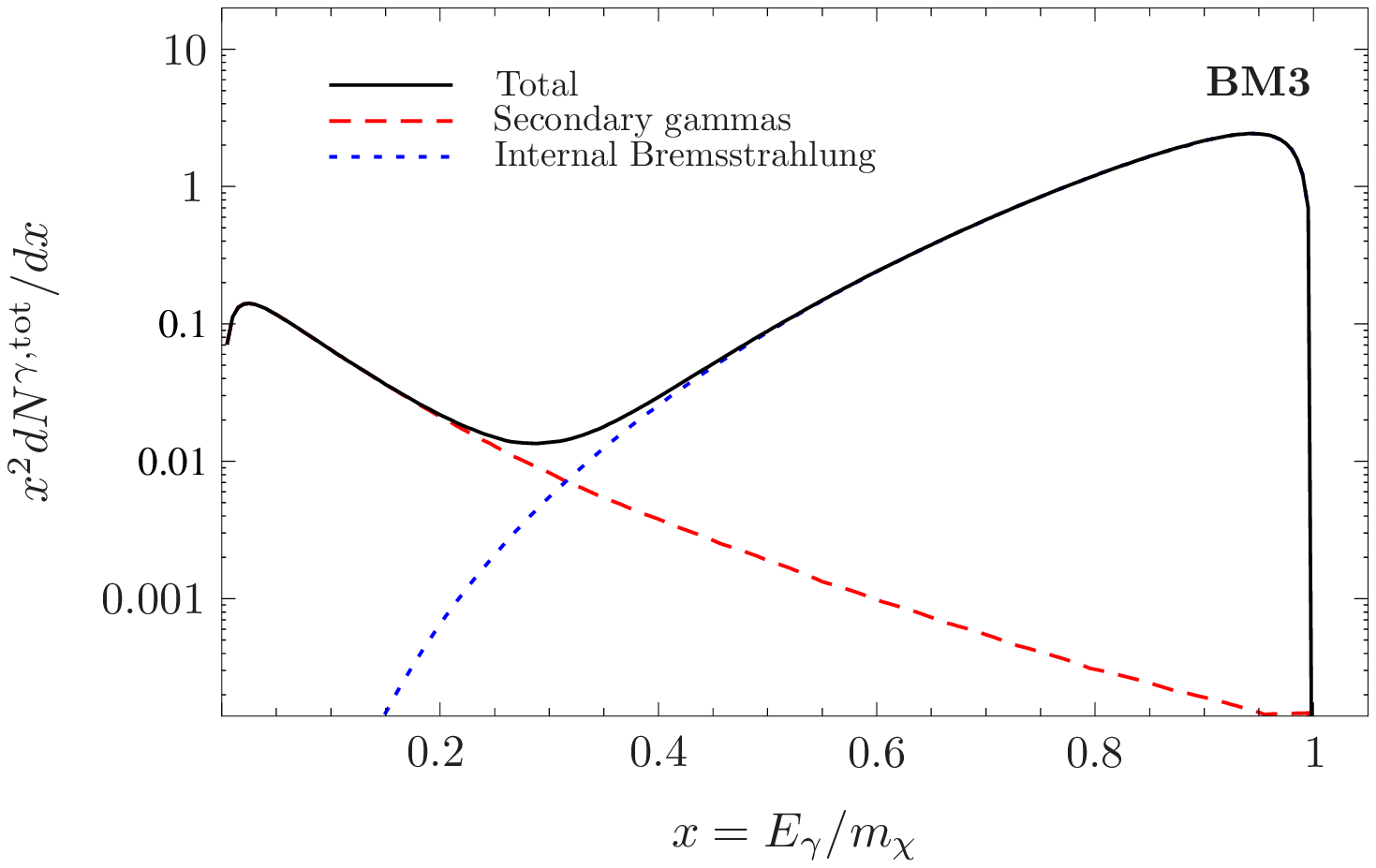}\\
\end{minipage}
\begin{minipage}[t]{0.02\textwidth}
\end{minipage}
\begin{minipage}[t]{0.49\textwidth}
\centering  
\includegraphics[width=\textwidth]{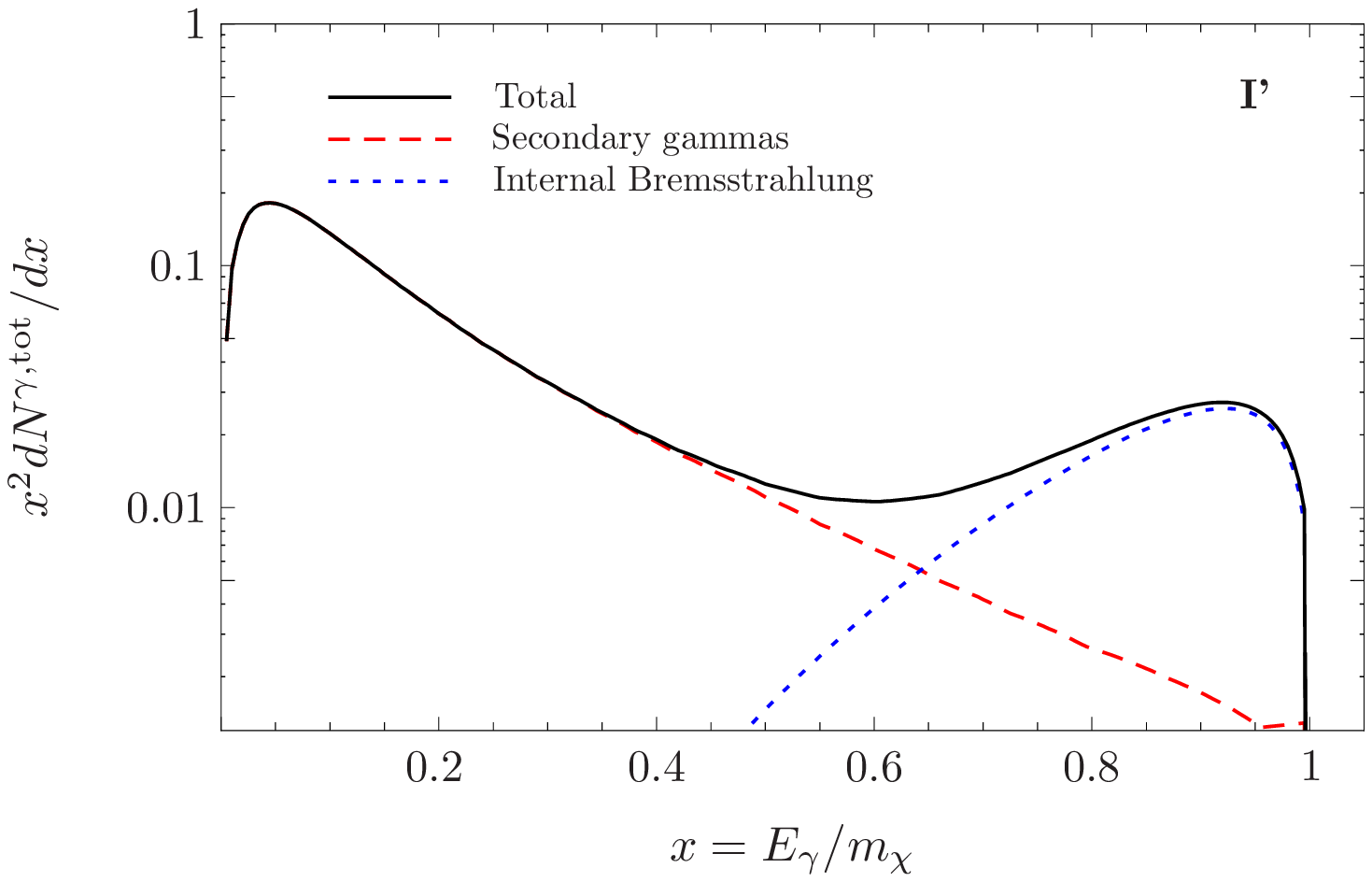}\\
\end{minipage}
\begin{minipage}[t]{0.49\textwidth}
\centering
\includegraphics[width=\textwidth]{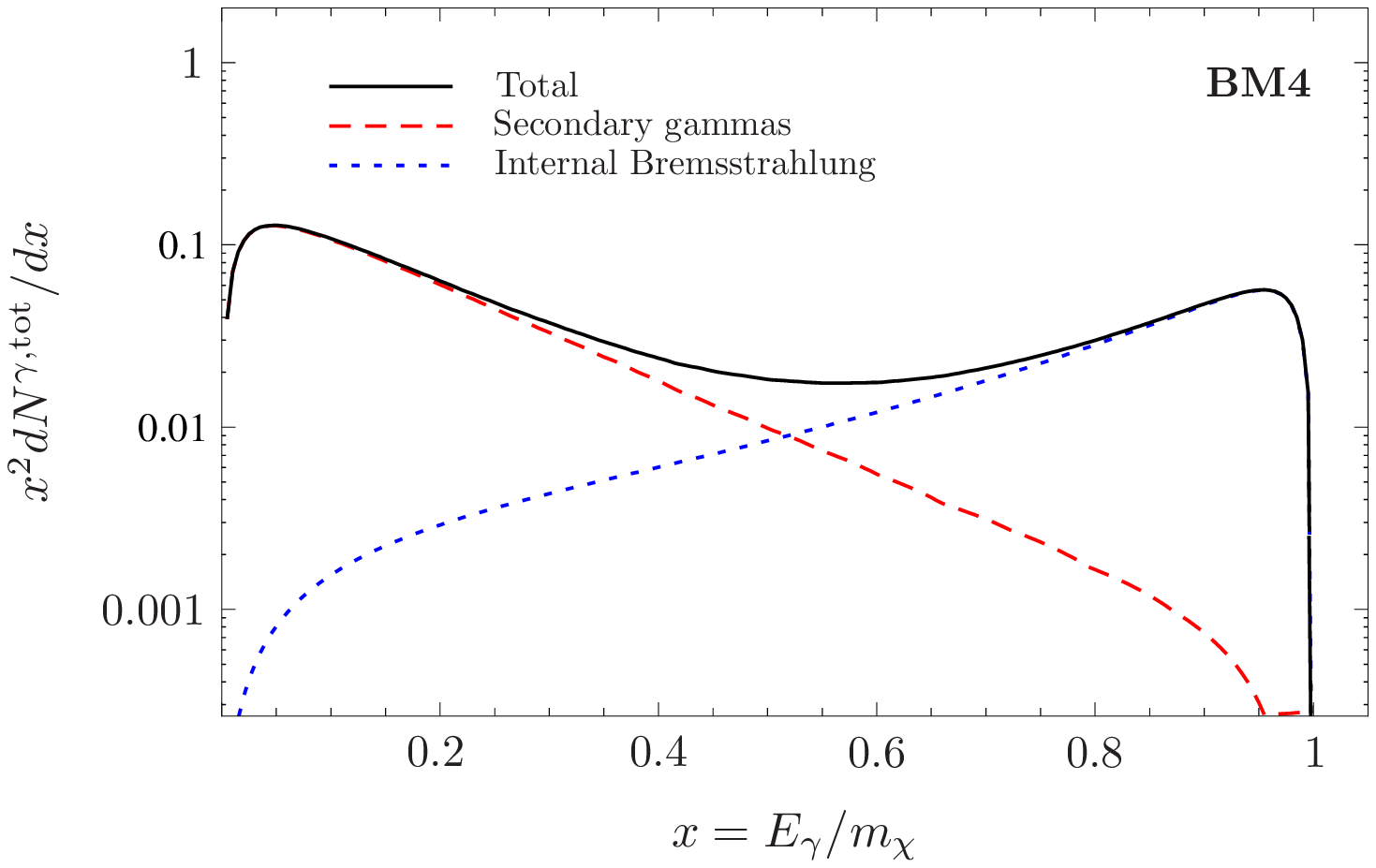}\\
\end{minipage}
\begin{minipage}[t]{0.02\textwidth}
\end{minipage}
\begin{minipage}[t]{0.49\textwidth}
\centering  
\includegraphics[width=\textwidth]{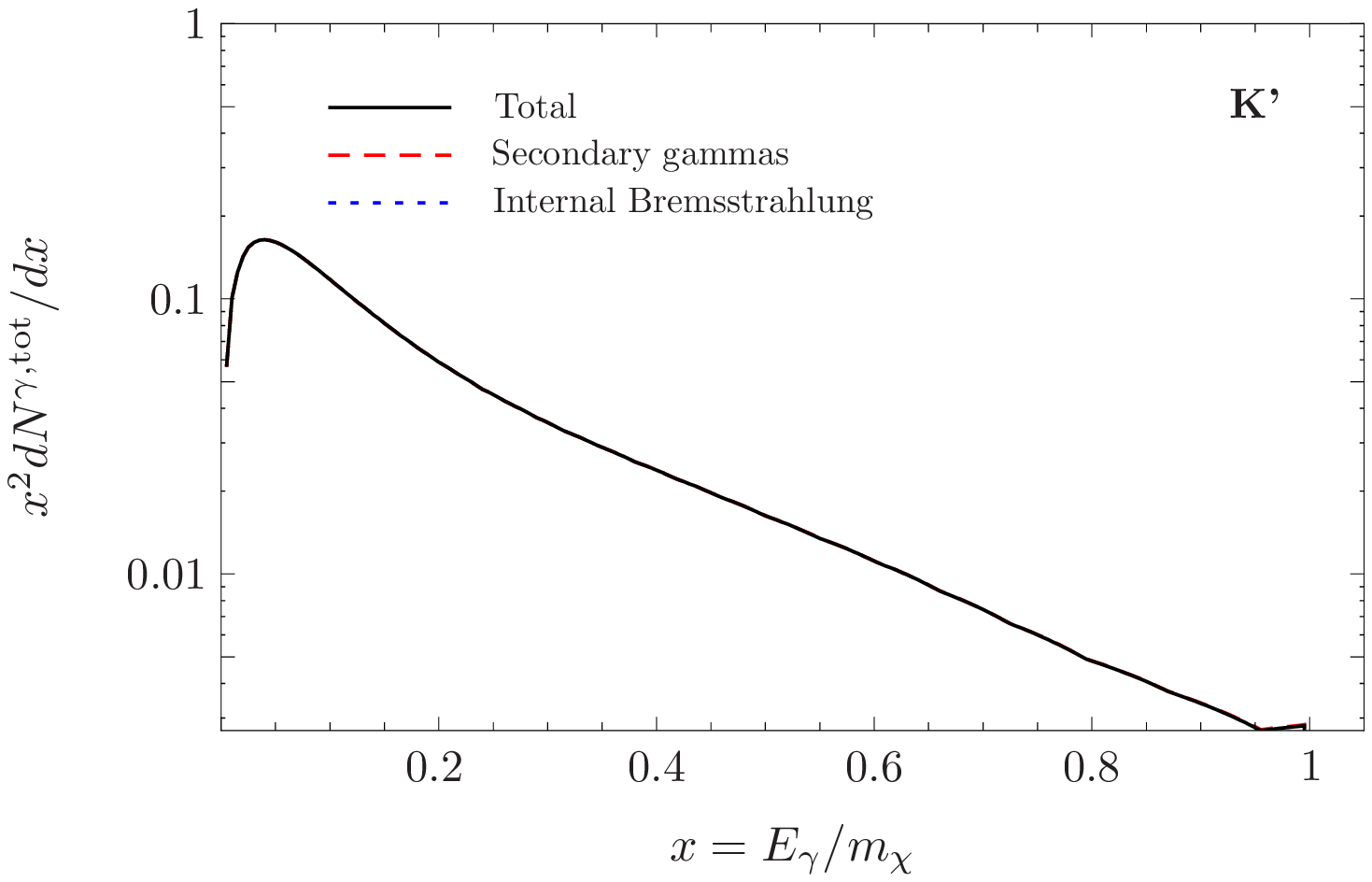}\\
\end{minipage}
\caption{The annihilation spectra in gamma rays for the cosmologically interesting regions of the mSUGRA parameter space, i.e.~the coannihilation region (BM3), the bulk region (I'), the focus point region (BM4) and the funnel region (K'). Line signals are not included. The benchmark points represent typical examples of these regions and are defined in  Refs.~\cite{IB_SUSY} (BM3,BM4) and \cite{Battaglia:2003ab} (I',K'), respectively.
\label{fig_spectra}}
\end{figure}

\begin{figure}[t]
\vspace*{-0.2cm}
\begin{minipage}[t]{0.61\textwidth}
\centering
\includegraphics[width=\textwidth]{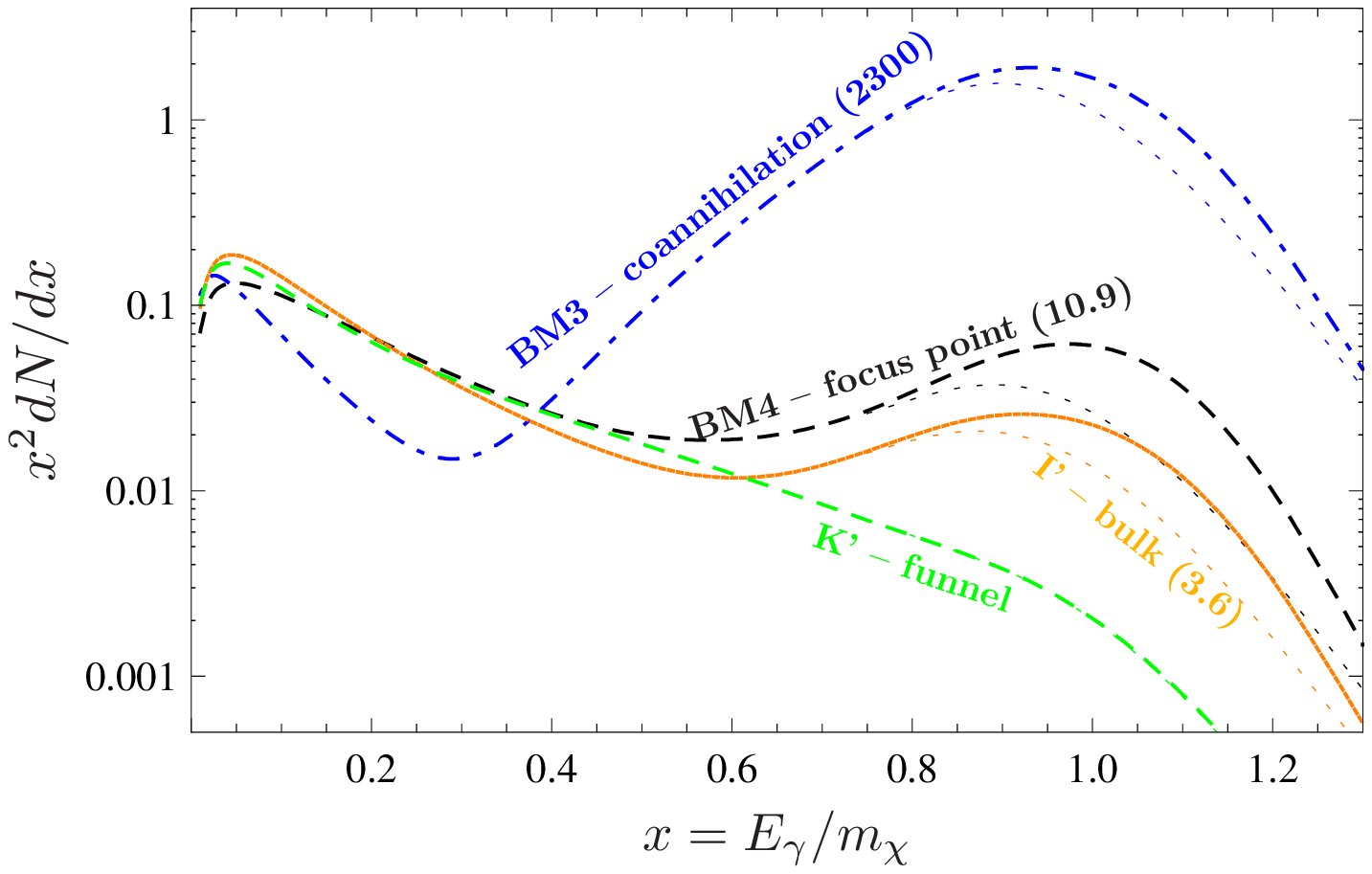}\\
\end{minipage}
\begin{minipage}[t]{0.02\textwidth}
\end{minipage}
\begin{minipage}[t]{0.37\textwidth}
\centering  
\vspace*{-5.98cm}
\includegraphics[width=\textwidth]{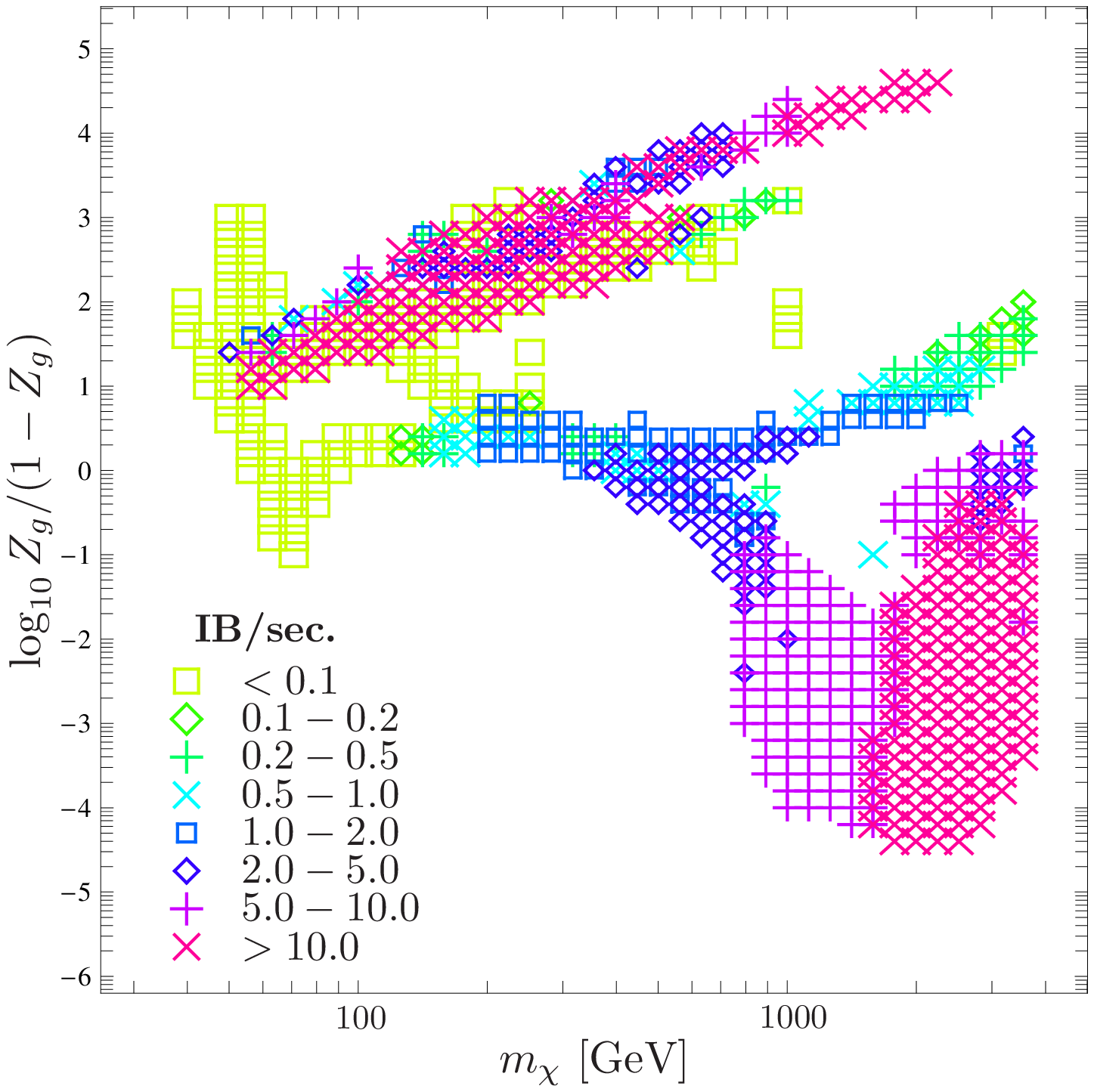}\\
\end{minipage}
\caption{
The spectra from Fig.~1
 are plotted together, as roughly seen by a detector with an energy resolution of 10\%. Here, the line contributions are also included (the dotted lines show the same spectra without them).
For each of these models, the IB enhancement is indicated in parenthesis, i.e.~the number of IB over secondary photons at energies $E_\gamma>0.6\,m_\chi$. 
For comparison, the right panel shows the result of a scan \cite{IB_SUSY} over the full mSUGRA parameter space, where this quantity is plotted as a function of the neutralino gaugino fraction $Z_g$  and  mass $m_\chi$. In the case of benchmark model K', both IB and line contributions are negligible.
\label{fig_spectra2}}
\end{figure}

Among indirect DM detection techniques, gamma rays play a pronounced role since they are usually more copiously produced than other possible messengers, directly trace the DM distribution as they propagate essentially absorption-free through the galaxy, and provide distinctive \emph{spectral signatures}. The last of these points is of particular importance to indirect searches because our knowledge about the distribution of DM, especially on small scales, is rather limited
and the \emph{amplitude} of the annihilation signal therefore usually subject to great uncertainties.
One can distinguish three different types of contributions to the gamma-ray spectrum: 
\begin{itemize}
\item At tree-level, WIMPs can annihilate into pairs of quarks, leptons, Higgs and (weak) gauge bosons, but not into photons. \emph{Secondary photons}, however, are produced in the hadronization and
further decay of the primary annihilation products, mainly through $\pi_0\rightarrow\gamma\gamma$. The result is a featureless spectrum with a rather soft cutoff at the DM particles' mass $m_\chi$, almost indistinguishable for the various 
possible annihilation
channels (with the exception of $\tau$-lepton final states). 

\item At $\mathcal{O}\left(\alpha_\mathrm{em}\right)$, another contribution has to be
included whenever charged annihilation products are present; \emph{internal bremsstrahlung} (IB), where an additional photon
appears in the final state. As  pointed out recently
\cite{FSR,IB_SUSY}, these photons generically dominate 
at high energies, i.e.~close to $m_\chi$, and thereby
add pronounced signatures to the spectrum; viz.~a very sharp cutoff at $m_\chi$ and bump-like features at
slightly smaller energies. 

\item Necessarily loop-suppressed, and thus only at $\mathcal{O}\left(\alpha_\mathrm{em}^2\right)$, 
\emph{monochromatic} $\gamma$
\emph{lines} result from the  annihilation of DM particles into two-body final states containing a photon \cite{lines}.
While providing a striking
experimental signature, these processes are usually subdominant (for a recent
analysis, see \cite{IB_SUSY});
examples of particularly strong line signals, however, exist \cite{idm}.

\end{itemize}

Fig.~\ref{fig_spectra} shows the annihilation spectra of neutralino DM in the case of minimal supergravity (mSUGRA), where one can single out four regions in the underlying parameter space that give the correct DM relic density (see, e.g., \cite{Battaglia:2003ab} for a discussion): after taking into account IB contributions, these spectra develop interesting, and evidently rather different features; only in the funnel region, where the mass of the pseudo-scalar Higgs boson is tuned such as to resonantly enhance neutralino annihilation, IB effects are negligible. Fig.~\ref{fig_spectra2} compares these spectra after smearing them with an energy resolution of 10\% (the design goal for the planned Cherenkov Telescope Array, CTA). Clearly, the spectra still remain well distinguishable; a detection would thus provide valuable information on the nature of the annihilating DM particles. The same figure also indicates the comparably small contribution from line signals and
states the ratio of IB over secondary photons at high energies, for the four benchmark models as well as for a full scan \cite{IB_SUSY} over the mSUGRA parameter space. This ratio can be as high as several orders of magnitude in the $\tilde \tau$-coannihilation region -- and therefore significantly improve the detectional prospects for these types of models, as shown recently in a study about DM annihilation signals from nearby dwarf galaxies \cite{draco}. 

IB effects in supersymmetry are dominated by contributions from photons radiated off charged virtual particles \cite{IB_SUSY}. Kaluza-Klein DM, another interesting example of WIMP DM, mainly annihilates into leptons; as a result, the spectrum takes the form typically expected from final state radiation \cite{UED,FSR} and is rather easily distinguishable from the spectra shown in Figs.~\ref{fig_spectra} and \ref{fig_spectra2} \cite{KKSUSYcomp}.

\section{Positrons}

The propagation of charged particles through the diffusive halo generally smoothens all features in the spectra of, e.g., positrons. Pronounced spectral signatures from DM annihilation as in the case of gamma rays can therefore more or less only be expected for exceptionally large branching ratios directly into $e^+e^-$ -- which, like in the case of Kaluza-Klein DM, leads to a very hard spectrum with an abrupt cutoff at $m_\chi$. Supersymmetric DM with its suppressed annihilation into leptons, on the other hand, generically produces rather soft positron spectra. Against this background, it was recently pointed out that radiative corrections in some cases can boost the annihilation of neutralinos into $e^+e^-\gamma$ final states sufficiently as to give a much harder positron spectrum than what is usually expected \cite{IBpos}. While slightly less pronounced than in the case of a direct annihilation into $e^+e^-$, the associated cutoff at $m_\chi$ would still provide a striking signature.

PAMELA \cite{pamela} has recently reported an unexpected excess in the positron flux at energies above around 1 GeV, rising with energy at a slope that agrees well with DM annihilation in the above-mentioned model. In order to explain the strength of the signal in this way, however, one would have to assume non-thermal DM production or a non-standard halo formation. Following Ref.~\cite{IBpos}, there have been quite a few attempts to interpret the observations in terms of DM annihilation; even more traditional astrophysical explanations exist. Since positron propagation is, furthermore, still bound to considerable uncertainties \cite{posprop}, one would in any case have to see a clear cutoff in the data before conclusively being able to infer a DM origin of the observed excess.

\section{Conclusions}

Radiative corrections generically alter DM annihilation spectra in gamma rays significantly, thereby adding distinct signatures that would not only provide smoking gun evidence for the particle nature of DM but
potentially also allow the discrimination between different DM candidates. In some cases the photon yield at high energies is boosted by several orders of magnitude; since air Cherenkov telescopes are most sensitive at these energies, this can lead to considerably better detectional prospects. Radiative corrections can, furthermore, lead to distinct spectral signatures in the positron flux in a way that would fit the excess recently reported by PAMELA; if the cutoff predicted at only slightly higher energies is confirmed by follow-up measurements, this would provide exciting insights into the nature of DM.

The radiative corrections reported here have been fully implemented in the most recent public release 5.0 of {\sf DarkSUSY} \cite{ds}.


\end{document}